\documentstyle[12pt,epsf,epsfig]{article}
\date{}
\newcommand{\be}{\begin{equation}}
\newcommand{\ee}{\end{equation}}
\newcommand{\bea}{\begin{eqnarray}}
\newcommand{\eea}{\end{eqnarray}}

\begin{document}
\begin{titlepage}

\title{Non-perturbative renormalization constants on the lattice from
flavour non-singlet \\ Ward identities}

\author{G.\ M.\ de Divitiis$^a$  and  R.\ Petronzio$^{b,}$\thanks{
Also at: Dipartimento di Fisica, Universit\`a di Roma {\em Tor Vergata} and 
 INFN, Sezione di Roma II, 
 Via della Ricerca Scientifica 1, 00133 Rome, Italy 
} \\
 \small $^a$ Department of Physics and Astronomy, 
 \small    University of Southampton, \\ 
 \small Southampton SO17 1BJ, UK \\
 \small $^b$ CERN, Theory Division, CH-1211 Gen\`eve 23,
      Switzerland\\
}

\maketitle

\begin{abstract}
	By imposing axial and vector Ward identities for flavour-non-singlet
	currents, we estimate in the quenched approximation
	the non-per\-tur\-ba\-ti\-ve values of 
	combinations of improvement coefficients, which appear in 
	the expansion around
	the massless case of the renormalization constants of
	axial, pseudoscalar, vector, scalar 
	non-singlet currents and of the renormalized mass.
	These coefficients are relevant for the completion 
	of the improvement programme to $O(a)$ of such operators. 
	The simulations are performed with a clover Wilson action
	non-perturbatively improved.
\vspace{0.8 cm}
\noindent
 \end{abstract}

\vfill
\begin{flushleft}
\begin{minipage}[t]{5. cm}
  { CERN-TH/97-274}\\
  { SHEP  97-24 }\\
October 1997
\end{minipage}
\end{flushleft}
\end{titlepage}

\section{Introduction}

	The programme of the improvement \cite{symanzik} 
	of the Wilson action 
	has been actively developed at the non-perturbative level 
	over the last years \cite{chiral_symm,
	non_pert_cSW_cA, non_pert_ren, non_pert_ZA_ZV_bV}. 
	At first, in the framework of the Schr\"odinger functional
	it was possible to determine  non perturbatively 
	the dependence upon the bare coupling constant 
	of the coefficient $c_{SW}$ \cite{non_pert_cSW_cA}
	of the clover term in the improved action \cite{SW}.

	Besides improving the action, the programme includes the
	improvement of the operators appearing in the
	correlation functions related to phenomenological 
	interesting quantities such as pseudoscalar meson decay constants 
	and the matrix element of the four-fermion
	operators of the weak effective Hamiltonian.

	In general the operator improvement consists of two parts:
	the mixing with higher-dimensional operators with 
	the same quantum numbers (in the literature the mixing coefficients 
	are called $c$) and the multiplication by a 
	suitable renormalization constant.
	The ultraviolet-finite renormalization constants can be expanded
	around the massless case:
\begin{equation}
Z_{\cal O} (m \neq 0) = Z_{\cal O} \cdot (1 + b_{\cal O} ma +...),
\label{eq:renorm_const}
\end{equation}
	where we have omitted corrections due to lattice artefacts of 
	order $a^2$ and higher.

	Some of these quantities have been calculated at the
	perturbative level 
	for axial ($A$), vector ($V$), pseudoscalar ($P$)
	and scalar($S$)
	currents as well as for the renormalized mass ($m$)
	\cite{pert_1, pert_2, pert_3}:
	non perturbative estimates are available for $Z_A$, $Z_V$,
	$b_V$ \cite{non_pert_ZA_ZV_bV} and for $c_A$ \cite{non_pert_cSW_cA}
	and $c_V$ \cite{non_pert_cV}. 
 
	In this letter, we present a  non-perturbative determination of
	the quantities:
\begin{equation}
b_A - b_P, \; \; \; \;   
b_V - b_S, \; \; \; \; 
b_m
\label{eq:b_coeff}
\end{equation}
	and of the ratios:
\begin{equation}
Z_m Z_P / Z_A, \; \; \; \; 
Z_m Z_S / Z_V  
\label{eq:Z_ratio}
\end{equation}
	from a set of axial and vector Ward identities.
\section{The method}
	The extraction of the $b$ coefficients and of the $Z$ ratios 
	is based on the following Ward identities:
\begin{eqnarray}
    \partial_\mu \langle {\mathbf A}^I_\mu(x) \Omega^\dagger (0) \rangle = &
  \!  2  m_{jk} & \! \! \! \langle {\mathbf P}(x) \Omega^\dagger (0) \rangle {\textstyle +O(a^2) }\nonumber\\
    \partial_\mu \langle {\mathbf V}^I_\mu(x) \Omega^\dagger (0) \rangle = &
  \!  \Delta m_{jk} & \! \! \! \langle {\mathbf S}(x) \Omega^\dagger (0) \rangle {\textstyle+O(a^2),}
\label{eq:ward_ident}
\end{eqnarray}
\noindent
	which, after $\vec{x}$ integration, become:
\begin{eqnarray}
    \partial_t \langle A^I_0(t) \Omega^\dagger (0) \rangle = &
  \!  2  m_{jk} & \! \! \! \langle P(t) \Omega^\dagger (0) \rangle {\textstyle +O(a^2) }\nonumber\\
    \partial_t \langle V^I_0(t) \Omega^\dagger (0) \rangle = &
  \!  \Delta m_{jk} & \! \! \! \langle S(t) \Omega^\dagger (0) \rangle {\textstyle+O(a^2).}
\label{eq:int_ward_ident}
\end{eqnarray}
	The suffix $I$ for the axial current indicates that 
	the current is improved by the appropriate mixing 
	with the pseudoscalar density multiplied by the
        coefficient $c_A$. The value for $c_A$ is taken from 
	its non-perturbative determination in ref.\cite{non_pert_cSW_cA}.
	For the vector current, 
	the contribution of the mixing with the tensor
	current (see ref.\cite{non_pert_cV} for non-perturbative 
	determination of 
	the mixing coefficient $c_V$) vanishes because of 
	the antisymmetry of tensor indices.

	The indices $j,k$ refer to the flavour content of the 
	bilinear operators, 
	which can be written as
\begin{equation}
      {\cal O} (t) = \sum_{\vec{x}} {\mathbf  O}  (x) = \sum_{\vec{x}} \bar \psi_j (x) \Gamma_{\cal O} \psi_k (x).
\end{equation}

	Equations (\ref{eq:ward_ident}) hold for any operator $\Omega$
	at any time different from $t$, reflecting the fact that 
	the W.I. are  identities among operators.
	For the axial W.I.\ we use $\Omega = P(0)$ and $\Omega = A_0(0)$ while
	for the vector W.I.\ we use $\Omega = S(0)$ and $\Omega = V_0(0)$; 
	in both cases the latter operator leads to much noisier results.

	The quantities $m_{jk}$ and $\Delta m_{jk}$ are 
	related respectively to the average and the difference  of 
	the renormalized mass, according to
\begin{eqnarray}
{m_{jk}} = m_{jk}^R 
\frac{Z_P}{Z_A} \frac{1 + b_P \overline{m_q a} }{1 + b_A \overline{m_q a} }  \nonumber\\
 \Delta {m_{jk}} = \Delta m_{jk}^R \frac{Z_S}{Z_V}\frac{1 + b_S \overline{m_q a} } 
{1 + b_V \overline{m_q a} } \label{eq:mdeltam}
\end{eqnarray}
$$ m_{jk}^R =  \frac{1}{2}(m_j^R + m_k^R), \; \; \; \Delta m_{jk}^R =  m_j^R - m_k^R, \nonumber
$$
	where $\overline{m_q a}$ is the average of the bare masses $j$ and $k$:
\begin{eqnarray}
 \overline{m_q a} =  \frac{1}{2}(m_{q_j}a + m_{q_k}a)\nonumber\\
 a m_{q_j}= \left( \frac{1}{2 \kappa_j} - \frac{1}{2 \kappa_c} \right)
\end{eqnarray}
	with $\kappa_c$ the critical value of the Wilson hopping parameter.
	The parameter $\kappa_c$ 
	is determined from the chiral extrapolation of the mass defined 
	through the axial Ward identities themselves. 

	The renormalization constants can be determined
	by replacing the ``unrenormalized current masses'' 
	$m_{jk}$ and $\Delta m_{jk}$ with the renormalized
	ones through the above equation, and then  the renormalized
	masses in terms of the bare masses:
\begin{equation}
m^R = Z_m m_q ( 1 + b_m m_q a).
\end{equation}
	Indeed, by including the lattice artefacts up to $O(a)$: 
\begin{eqnarray} {
m_{jk} a = \frac{Z_P Z_m}{Z_A} \left(\overline{m_q a}  - (b_A - b_P) 
(\overline{m_q a})^2
 + b_m \overline{(m_q a)^2}\right)   }\nonumber\\
{\Delta m_{jk}a = \frac{Z_S Z_m}{Z_V} (m_{q_j}a - m_{q_k}a )\left(1 + 
( 2 b_m - (b_V - b_S)) \overline{m_q a} \right)}
\protect\label{eq:mainequation}
\end{eqnarray}

	From a fit of the bare mass dependence of these results 
	we can determine non-perturbatively the combinations 
	in eqs. (\ref{eq:b_coeff},\ref{eq:Z_ratio}).

	Within the approximation
	used in the above formulae, we can derive an expression for
	$(b_A - b_P)$:\\
\bea
m_{jk}a-\frac{( m_{jj}a+m_{kk}a )}{2} = \frac{Z_P Z_m}{Z_A}\frac{1}{4} (b_A - b_P)
( m_{q_i}-m_{q_j} )^2 a^2
\protect\label{eq:bAP}
\eea
	which can be used to determine a value for the combination
	independently from the knowledge of the critical value of $\kappa$.
	We have compared such a determination of $b_A-b_P$ with the one
	coming from a global fit to expression eq.(\ref{eq:mainequation})
	and used it as a sensitive check of our estimate of $\kappa_c$.

	The presence of flavour-non-diagonal currents is
	important for the fit of the axial W.I. and essential 
	for the vector ones.
	We want to point out that $m_{jk} a$ depends either upon 
	$(\overline{m_q a})^2$ or upon $\overline{(m_q a)^2}$,
	which are different variables if the flavours are not 
	degenerate, making then possible to disentangle the 
	coefficient $b_m$ from the combination $b_A - b_P$.	

	Our fits are to the dependence upon the quark mass, 
	and order $a^2$ corrections linear in the quark mass can in 
	general fake the extraction of the coefficients.
	The simple lattice discretization of the time derivative
	$\frac{1}{2a}\scriptstyle (f(t+a)-f(t-a))$ has an error 
	of the order of $f^{(3)} a^2$, which becomes $f^{(1)} M^2a^2$
 	when a single state of mass $M$ dominates 
	the correlation function $f$. In the pseudoscalar channel, 
	chiral symmetry makes this term
	proportional to the quark mass.
	In other channels, the mass $M$ acquires a non-zero value
 	when the quark masses vanish.
	The quantity $(Ma)^2$ still contains a term that is linear in the 
	quark mass but not in the lattice spacing.
	In order to minimize such effects
	we have used the lattice discretization of the time derivative
	$a  \partial_t$
	correct up to term $f^{(5)}a^5$.
	While for the pseudoscalar case with spontaneous symmetry
	breaking the improvement to the
	fifth order of the derivative removes these fake linear terms
	in the quark mass, for the vector current a linear term
	survives at any finite order $n$, although suppressed by a 
	coefficient of order $1/(n-1)!$. We have checked that improving
	the derivative to the next order does not change our results 
	beyond their accuracy.
	
	We cannot exclude the presence of other lattice artefacts 
   	in the ratio of matrix elements
	formally of order $a^2$ but linear in the quark mass.
	We have checked in the case of the axial current that
	the results are stable with respect to the choice of the operators
	and we interpret this as a sign for the absence of large
	extra terms of order $m \Lambda a^2$.

	This method of computing the combinations
	in eqs. (\ref{eq:b_coeff},\ref{eq:Z_ratio})
	is valid in the quenched approximation.
	The presence of dynamical quark loops would introduce 
	an additional sea quark mass dependence, which would involve 
	flavours different from those in the currents.

\section{The numerical results}
\begin{table*}[t]
\setlength{\tabcolsep}{1.pc}
\newlength{\digitwidth} \settowidth{\digitwidth}{\rm 0}
\catcode`?=\active \def?{\kern\digitwidth}
\begin{tabular*}{\textwidth}{@{}l@{\extracolsep{\fill}} l l l l l}
\hline
 $L^3 \; T$  &  $16^3 48$  &  $16^3 32$  &  $16^3 32$  &  $16^3 32$  \\
 $\beta$       &  $6.2$      &  $6.8$      &  $8.0$     &  $12.0$     \\
 \# confs       &  50         &  50         &  80         &  80     \\
\hline
    &  &  &  &         \\
$\kappa$  &  0.124    &  0.124967  &  0.129382  &  0.126299     \\
          &  0.1275   &  0.127517  &  0.130055  &  0.126941     \\
          &  0.1295   &  0.128831  &  0.130736  &  0.127589     \\
          &  0.132    &  0.131198  &  0.131078  &  0.127915     \\
          &  0.13326  &  0.132589  &  0.131423  &  0.128243     \\
          &  0.13362  &  0.132942  &  0.131700  &  0.128507     \\
          &  0.134    &  0.133296  &  0.131908  &  0.128705     \\
          &  0.1345   &  0.133796  &  0.132222  &  0.129004     \\
          &  0.135    &  0.134371  &  0.132467  &  0.129237     \\
          &  0.13535  &  0.134660  &  0.132749  &  0.129505     \\
    &  &  &  &         \\
$\kappa_c$  &  0.13578(2)  &  0.13511(1)  &  0.13318(1)  &  0.129915(8)     \\
    &  &  &  &         \\
\hline
\end{tabular*}
\caption{The values of $\kappa$ used in our simulations at various $\beta$}
\vspace{0.3 cm}
\protect\label{tab:1}
\end{table*}
	We have performed several simulations at different values of
	$\beta$, in order to derive the coupling constant dependence of the
	quantities in eqs. (\ref{eq:b_coeff},\ref{eq:Z_ratio}).
	The values of $\beta$ used in the simulations 
	and the corresponding volumes are collected in Table \ref{tab:1},
	with a list of the values of $\kappa$ and of our best estimate of
	the critical $\kappa$ obtained from the W.I. themselves.
	The values of $\kappa_c$ are well compatible within errors
	with those of ref. \cite{non_pert_cSW_cA}.
	The variation of our results under a change of $\kappa_c$ within the 
	quoted error is smaller than the accuracy by which we can extract the
	non-perturbative quantities from our fit.

	Simulations are performed with an updating sequence
	made by a standard heat-bath followed by 3 over-relaxation steps.
	The improved fermion propagator is calculated every 1000 gauge update
	using a stabilized biconjugate algorithm.
	For our runs we have used the 25 Gflops machine of the APE series,
	made of 512 nodes working in SIMD (Single Instruction
	Multiple Data) mode.

	Each propagator was  summed over the space volume
	distributed to the single node ($3\times 3 \times 2$)
	and stored on disk.
	The use of these
	propagators leads to correlation functions that 
	contain the correct local-gauge-invariant terms and
	other non-local, gauge-non-invariant terms that 
	go to zero after summing over the gauge configurations.
	We have explicitly checked that with our statistics 
	the residual noise due to imperfect cancellation of 
	the gauge-non-invariant terms is much below the statistical 
	fluctuations. Storing fermion propagators allows for
	an off-line calculation of all flavour-non-singlet correlations.
 
	The W.I. are satisfied separately at each time:
	after some initial time, up to which higher-order 
	lattice artefacts still dominate,  $m_{jk}$ and
	$\Delta m_{jk}$ show a plateau. 
 	At $\beta = 6.2$ we run two temporal extensions ($32$ and 
	$48$) in order to
	monitor the stability of the plateau.
	We have used two methods of analysis:
	either we first average the result over the time interval 
	of the plateau and then perform a fit, or
	we first perform a fit at each time value inside the plateau
	and then average the results of the fit at different times.
	The two procedures give very similar results. 

	The choice of the spectator operator $\Omega$ 
	affects the statistical error of the final results. 
	We have found that the pseudoscalar and the scalar densities give 
	the best results for the axial and the vector case
	respectively.

	We perform a fit of $2 m_{jk}$ with the function 
	(here all masses are in lattice spacing units): 
\newfont{\smallmath}{cmr10 scaled 500}
{\smallmath\begin{displaymath} 
a_1 (m_{q_i}+m_{q_j}) + a_2 (m_{q_i}^2+m_{q_j}^2) + a_3 (m_{q_i}+m_{q_j})^2 +
a_4 (m_{q_i}^3+m_{q_j}^3) + a_5 m_{q_i} m_{q_j} (m_{q_i}+m_{q_j})
\end{displaymath}}
	The first three coefficients
	of the fit are related to the renormalization constants as follows:
	$a_1 = Z_m Z_P / Z_A$; $a_2 /a_1 = b_m$; $a_3 /a_1 = - (b_A - b_P)/2$.
	The last two coefficients in the fit can be introduced
	to parametrize order-$a^2$ corrections compatible with the
	flavour exchange symmetry of the Ward identity.

	For the quantity  $\Delta m_{jk}$ we perform the fit with: 
{\smallmath\begin{displaymath}
 v_1 (m_{q_i}-m_{q_j}) + v_2 (m_{q_i}^2-m_{q_j}^2) + v_3 (m_{q_i}^3-m_{q_j}^3) + v_4 m_{q_i} m_{q_j} (m_{q_i}-m_{q_j})
\end{displaymath} }
	where
	$v_1 = Z_m Z_S / Z_V$; 
	$v_2/v_1 = b_m - (b_V-b_P)/2$. 
	As before, the extra coefficients $v_3$ and $v_4$  parametrize 
	the possible order-$a^2$ corrections.

	Our results normally refer to the fit with three parameters
	for the axial case and two for the vector case. 
	Increasing the number of parameters in general does not 
	improve the value of $\chi^2$ much while it considerably increases 
	the error.
	The results are compatible with the lower parameter fit,
	with the exception of the determination of $b_V$, which comes 
	systematically higher with the four-parameter fit. We have
	included this effect in the corresponding error.
\begin{table*}[t]
\begin{tabular*}{\textwidth}{@{}l@{\extracolsep{\fill}} r r r r r}
\hline
 $L^3 \; T$  &  $16^3 48$  &  $16^3 32$  &  $16^3 32$  &  $16^3 32$  \\
 $\beta$       &  $6.2$      &  $6.8$      &  $8.0$     &  $12.0$     \\
 \# confs       &  50         &  50         &  80         &  80     \\
\hline
    &  &  &  &         \\
  $Z_m Z_P / Z_A$          & 1.09(1) & 1.08(1)&  1.08(1)& 1.060(6)     \\
  $Z_m Z_S / Z_V$          & 1.24(2) & 1.21(1)& 1.142(4)& 1.080(5)     \\
  $b_A - b_P$              & 0.15(2) & 0.10(2)&  0.06(2)& 0.04(2)      \\
  $b_m$                    & $-0.62(3)$ & $-0.58(3)$&  $-0.57(3)$& $-0.53(2)$   \\
  $b_m-(b_V - b_S)/2 $     & $-0.69(4)$ & $-0.63(4)$&  $-0.54(3)$& $-0.52(2)$   \\
    &  &  &  &         \\
\hline
\end{tabular*}
\caption{The results of our calculations}
\vspace{0.3 cm}
\protect\label{tab:2}
\end{table*}

	Table \ref{tab:2} contains the main results, the values for
	the various renormalization parameters at different 
	$\beta$ and volumes. The  $\beta = 6.2$ results on the
	smaller temporal extension are fully compatible.\\
	With the values of the fit we can check if the renormalized 
	W.I. depend only upon the sum (for the axial) or 
	the difference (for the vector) of the renormalized masses.

\begin{figure}
\centerline{\epsfig{figure=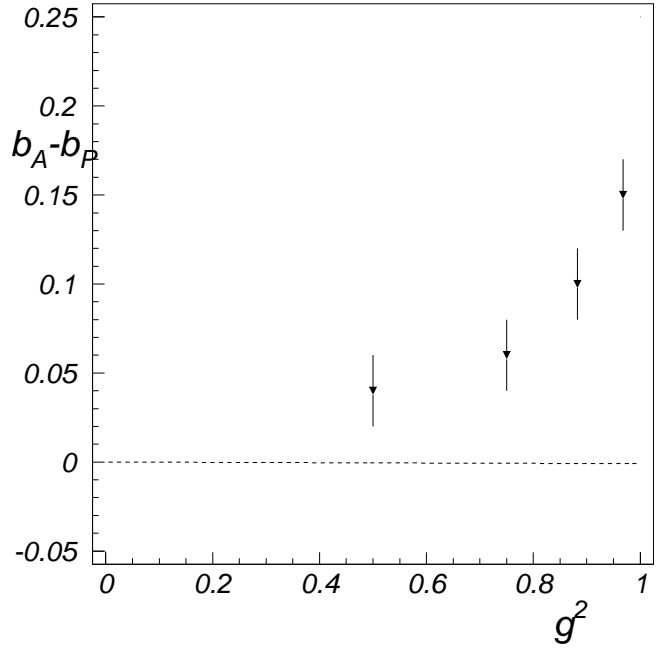,width=0.7\linewidth}}
\caption{
The non-perturbative result for $b_A - b_P$. The perturbative result
 of $O(g^2)$ is negligible on this scale. }
\protect\label{fig:b_APcoeff}
\end{figure}

\begin{figure}
\centerline{\epsfig{figure=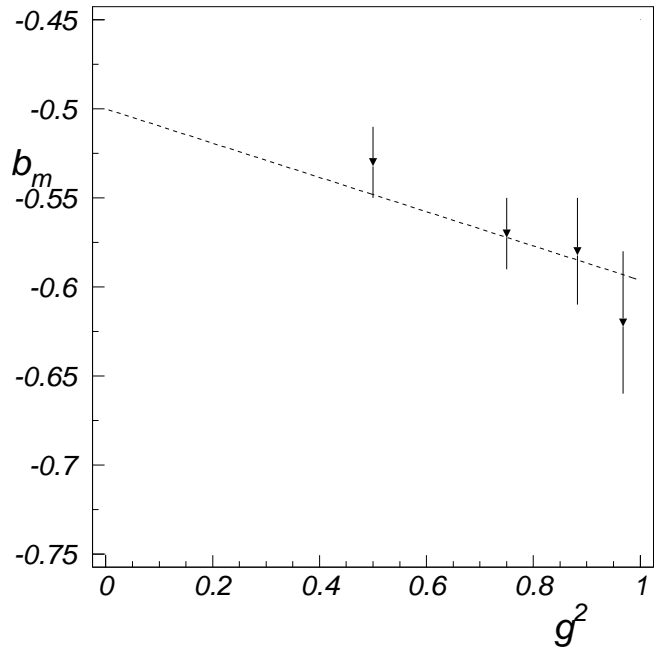,width=0.7\linewidth}} 
\caption{The non-perturbative estimate of $b_m$ is compared with the
perturbative result.}  
\protect\label{fig:b_mcoeff}
\end{figure}

\begin{figure}
\centerline{\epsfig{figure=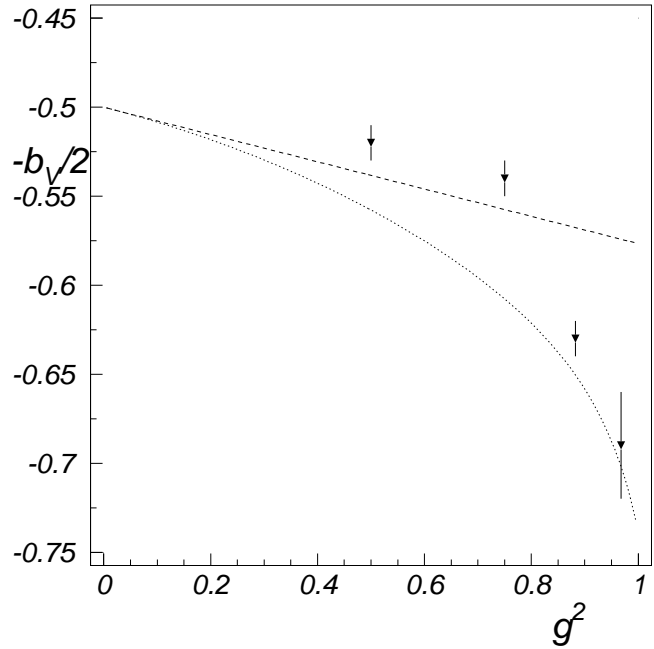,width=0.7\linewidth}}
\caption{
The non-perturbative result for $b_V$ is compared, after
using L\" uscher's relation, with the one of ref. [6] (dotted curve)
and with the perturbative result (dashed curve).}
\protect\label{fig:b_Vcoeff}
\end{figure}

\begin{figure}
\centerline{\epsfig{figure=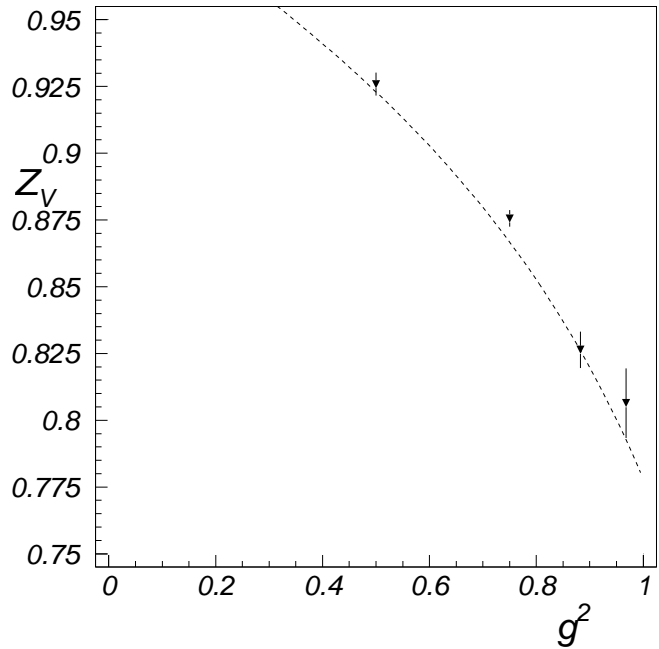,width=0.7\linewidth}}
\caption{
The non-perturbative result for $Z_V$ is compared, after
using L\" uscher's relation, with the one of ref. [6].
}
\protect\label{fig:Z_Vcoeff}
\end{figure}

	The renormalized masses and currents manage to bring on the same
	straight line
	points that appear misaligned and on a curved line
	for the bare quantities.
	For large values of the masses, higher-order terms enter 
	the game and produce again a misalignment of the 
	corresponding points.

	The fits in general do not include all $\kappa$ values; 
	we exclude the heavier masses until the stability of the 
	results is reached. 

	Our results, when compared with those available from 
	perturbation theory, 
	show that higher-order terms seem to dominate for the differences
	$b_A - b_P$, which is very small at one-loop order 
	(see fig. \ref{fig:b_APcoeff}),
	while  for $b_m$ the presence of sizeable terms of order
	$g^2$ makes the effect of $g^4$ terms less prominent.
	Indeed, our results are not far from lowest-order 
	perturbation theory for this case (see fig. \ref{fig:b_mcoeff}).\\
	For $b_V-b_S+2b_m$, there is an argument due to Martin L\" uscher
	\cite{privcomm}
	relating $b_m$ with $b_S$ and $Z_S$ with $Z_m$
	in the quenched approximation:
	$2b_m -b_S = 0$ and $Z_S Z_m=1$, which
    	implies that from the W.I. for the vector current 
	we actually obtain
	$b_V$ and $Z_V$. The comparison with those
	of ref. \cite{non_pert_ZA_ZV_bV} is
	shown in figs. \ref{fig:b_Vcoeff} and \ref{fig:Z_Vcoeff}:
	while for $Z_V$ there is a
	perfect agreement, for $b_V$ we are generally closer
	to the perturbative result.
	Our large errors are mainly systematic and reflect
	the instability of a four-parameter fit.
	Running at lower quark masses could reduce the discrepancy
	which might also be due to residual order-$a^2$ lattice artefacts.

\vspace{0.5 cm}

	The use of axial and vector Ward identities with flavour-non-singlet
 	currents allows the determination 
	in the quenched approximation of various non-perturbative
	renormalization constants. 
	The calculation that we have presented could be refined 
	by using the Schr\"odinger functional method
	which would allow a safe investigation of the very 
	low quark mass region.

\vspace{1.0 cm}

ACKNOWLEDGEMENTS. We thank M. Masetti for his collaboration to the first
part of this work. The program for the inversion of the fermion
propagator with the biconjugate gradient algorithm was written for the 
APE machine in Tor Vergata by A. Cucchieri and T. Mendes. 
We thank M. L\" uscher for various
interesting remarks and for stimulating discussions.

\newpage

\def\NPB #1 #2 #3 {Nucl.~Phys.~{\bf#1} (#2)\ #3}
\def\NPBproc #1 #2 #3 {Nucl.~Phys.~B (proc. Suppl.) {\bf#1} (#2)\ #3}
\def\PRD #1 #2 #3 {Phys.~Rev.~{\bf#1} (#2)\ #3}
\def\PLB #1 #2 #3 {Phys.~Lett.~{\bf#1} (#2)\ #3}
\def\PRL #1 #2 #3 {Phys.~Rev.~Lett.~{\bf#1} (#2)\ #3}
\def\PR  #1 #2 #3 {Phys.~Rep.~{\bf#1} (#2)\ #3}

\def\etal{{\it et al.}}
\def\ibid{{\it ibid}.}


\begin{thebibliography}{9}
\bibitem{symanzik}
   K. Symanzik, \\
   \NPB B226 1983 187, 205
\bibitem{chiral_symm}
   M. L\" uscher, S. Sint, R. Sommer and P. Weisz,\\
   \NPB B478 1996 365
\bibitem{non_pert_cSW_cA}
   M. L\" uscher, S. Sint, R. Sommer, P. Weisz and U. Wolff, \\
   \NPB B491 1997 323
\bibitem{non_pert_ren}
K. Jansen, C. Liu, M. L\" uscher, H. Simma, S. Sint, R. Sommer,\\ P. Weisz
and U. Wolff, \\
\PLB B372 1996 275
\bibitem{non_pert_ZA_ZV_bV}
M. L\" uscher, S. Sint, R. Sommer and H. Wittig, \\
\NPB B491 1997 344
\bibitem{SW}
   B. Sheikholeslami and R. Wohlert, \\
   \NPB B259 1985 572
\bibitem{pert_1}
M. L\" uscher and P. Weisz,\\
\NPB B479 1996 429
\bibitem{pert_2}
S. Sint and P. Weisz,\\
{\it Further results on $O(a)$ improved lattice QCD the one-loop order},\\ 
hep-lat/9704001
\bibitem{pert_3}
S. Sint,  {\it  Proceedings of the International Symposium on \\Lattice Field Theory, 21-27 July 1997, Edimburgh, U.K.}
\bibitem{non_pert_cV}
M. Guagnelli, {\it  Proceedings of the International Symposium on \\Lattice Field Theory, 21-27 July 1997, Edimburgh, U.K.}
\bibitem{privcomm}
M. L\" uscher, private communication


\end{thebibliography}
\end{document}